\documentclass[preprint,12pt]{elsarticle}
\usepackage[utf8]{inputenc}
\usepackage{amssymb}
\usepackage{subfig}
\usepackage{graphicx} 

\journal{Energy}

\begin{document}
\begin{frontmatter}


\title{Brazil electricity needs in 2030: trends and challenges}


\author{Marcos Paulo Belançon$^{a}$}
\ead{marcosbelancon@utfpr.edu.br}
\address[$^a$]{Universidade Tecnológica Federal do Paraná (UTFPR), Câmpus Pato Branco, Brazil}

\begin{abstract}
The demand for electricity and the need to replace fossil fuels by renewables have been growing steadily, and this transition will have significant implications to our world that are only beginning to be understood. Brazil is one important example of a big economy where the electricity is already supplied by renewables, such as hydro, wind and biomass-fired thermal power. In this work we investigated the electricity load curves in the last 20 years in Brazil, and four different scenarios for 2030 are proposed in order to evaluate the impact of increasing renewables in the national grid, at an hourly basis. The analysis shows that growing electricity demand and the expected reduction in the hydropower share will significantly increase the reliability of the national grid, due to higher peak load and also due to the intermittency of Solar and Wind. Without any gigawatt scale hydropower projected for the near future, increasing the share of these renewables should push hydropower to operate hundreds of hours every year above typical peak power levels experienced in the past. In order to avoid or reduce the threat related to this trend one of our scenarios suggests that solar water heaters could be massively deployed in Brazil, what would positively impact the system reliability by reducing the electricity demand mostly at peak loads during early evenings.

\end{abstract}
\end{frontmatter}
\section{Introduction}

Brazil has more than 200 million inhabitants and is the fifth biggest country in the world by land area. By several parameters, such as gross domestic product (GDP) and human development index (HDI) the country represents the world average and it can be seen as a sample of the entire world. 

The growing need to decarbonize our energy system worldwide and mitigate climate change \cite{Chu2012,Clack2017,Gils2017,Peters2017a,Oberthur2016,Hulme2016} makes noteworthy the unique proportion of renewables in Brazil's primary energy \cite{BEN2019}, which is three times superior to the world average (45 \% against only 15\%). This makes Brazil the ``greenest'' among the biggest economies of the world: its Carbon intensity is about 0.15 $kgCO_2/US\$_{ppp}$, which is lower than in Europe (0.18), United States (0.29), and China (0.47).

Hydropower is by far the most important source of electricity and until a few decades ago this single-source supplied more than 95\% of the demand. But since the 2000's concerns \cite{Almeida2001} about the challenges of expanding electricity production without relying on fossil fuels are growing. In 2001 drought conditions \cite{Jardini2002,Hunt.2018} and delay in generation investments resulted in one of the worst energy crises of the modern times, which sounded the alarm to enhance energy security by diversifying its electricity sources.

Despite some huge investments in new hydro capacity \cite{Fearnside2017,Humood2017} since the 2000's, in 2014 a new energy crisis was triggered by constrained hydropower due to low precipitation levels \cite{MarengoOrsini2018} and minor factors such as mismanagement of the regularization reservoirs \cite{Hunt.2018}. The expansion in nominal hydropower capacity in the last 10 years was not translated into higher hydropower generation. The production has peaked in 2011 (449 TWh) and remained about 10\% below this level since then, mainly due to droughts in the southeast of the country and the head loss in several strategic reservoirs in this area, which has as consequence also impacted in the hydropower revenues \cite{DeQueiroz2019}.

Other renewables have significantly expanded in Brazil, namely biomass-fired thermal and wind. However, gas-fired thermal power is on the rise and is expected to be the second most important source of electricity in the coming years, pushing up greenhouse gas (GHG) emissions from power generation \cite{DeFaria2017}. This trend is the result of the increasing power demand and the intermittent nature of Wind and Solar, which cannot meet the load curves \cite{Diuana2019}.

Briefly one may enumerate some relevant questions about the future of electricity production in Brazil:
\begin{enumerate}
\item Is there a fossil-fuel-free pathway to enhance the reliability of the supply?
\item How much intermittency of solar and wind can be fulfilled by hydropower?
\item Among the generation technologies available, which one should be a priority?
\end{enumerate}

To address these questions, in this work we have analyzed the load curves of the Brazilian national grid, covering the period between 1999 and 2018. Generalized Additive Models were used to better identify the trends. A 24\% increase in electricity demand for 2030 was considered, and we have evaluated four different scenarios for how this demand could be fulfilled, investigating its impacts and consequences.

\section{Methods}

\begin{figure}[!h]
\centering
\includegraphics[scale=0.4,clip]{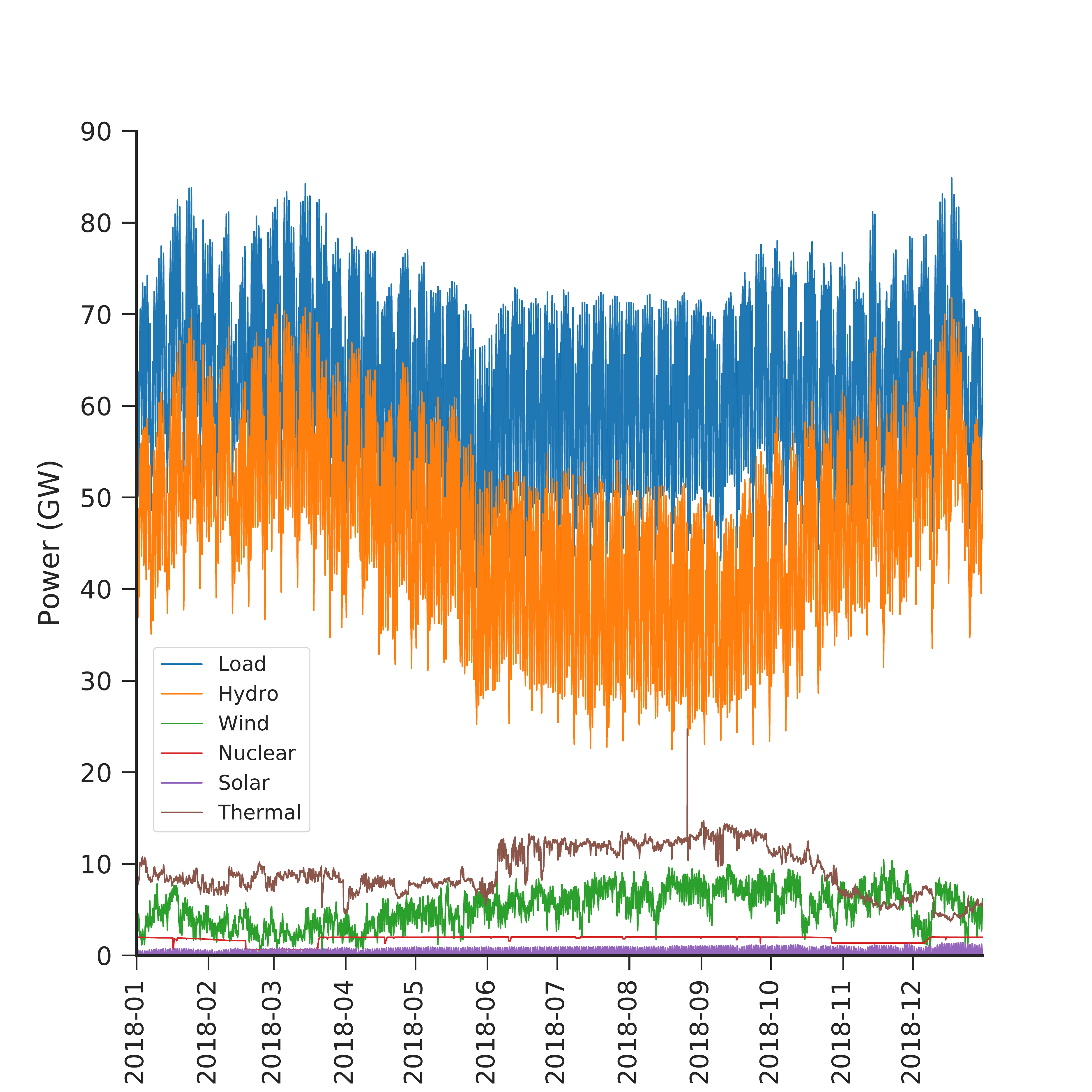}%
\caption{Evolution of power consumption in Brazil with hourly resolution.}
\label{fig2018}
\end{figure}

Data on the national system grid load and generation by source over the period 1999-2018 were investigated, with hourly resolution. As an example, the data for the year 2018 is shown in figure \ref{fig2018}. Besides some analyses direct performed with the raw data, Generalized Additive Models\cite{Hill1986} were used to better identify trends.

In the last two decades, electricity consumption has grown as fast as 4\% per year in Brazil, though since 2014 it has averaged only 1\%. In this context, we found reasonable to assume a growth rate of 2\% for the period 2018-2030, which is similar to the low growth scenario from ``Empresa Brasileira de Pesquisa Energética'' (PNE2030). At such a rate in 2030 Brazil's annual consumption of electricity should increase by 137 TWh and reach 689 TWh, against 551.8 TWh in 2018. 

The next step in our study was to project a load curve demand for 2030. Contour plots of the load power against year and day of the week, as we are going to show, indicated that in the last two decades the load peak has been shifted from early evenings across the entire year to afternoons in February/March. As there is no sign that this trend will be reversed in the coming years, we fixed our targeted load curve for 2030 as 1.24 (689TWh/551.8TWh) times the load curve in 2018.

Brazil has today about 85 GW of hydroelectric dams and 12 GW of run-of-river hydropower \cite{Dranka2018a}, in such a way that hydro is the only source with a nominal capacity higher than the annual load peak. As no major new gigawatt-scale plant is expected to come online until 2030, we decided to analyze scenarios where total hydropower production per year do not exceed the 2018 level, in such way that the new demand expected for 2030 needs to be supplied by thermal, wind, nuclear and solar power or partially replaced by solar water heaters. In this way, one can evaluate how the expansion of these sources will affect the hydropower during an entire year, at hourly basis.
 
For example, when considering a two-times increase in wind energy production we used the wind generation curve shown in figure \ref{fig2018} multiplied by two, enabling the possibility to access the effects of this source could bring to the system. Even though such an approach is quite simple, as one year has about 8760 hours, this method provides some statistical insight about daily and seasonal trends when all different sources are combined.

\subsection{Scenarios}

In order to build projections, we need to take into account some fundamental aspects of Brazil's electricity mix. While Wind and solar power are widely known by its intermittency characteristic, thermal power in Brazil was historically provided by biomass, which is seasonal dependent on the climate. Though biomass has been surpassed by Gas-fired thermal power in the last few years \cite{Almeida2018}. Only two nuclear reactors are running in Brazil and the share of nuclear is not expected to change significantly in the next decade, since there is only one reactor in the construction phase, which was halted in 2014.

In this way, we are analyzing four scenarios, namely 2030x, where x is a,b,c, and d, which are not proposed to compare its feasibility, cost or to consider the construction of massive infrastructure \cite{DeBarbosa2017}. Indeed they are used as tools to identify the challenges the national grid is going to face and the ability of hydropower to keep filling the gap between the supply of intermittent sources and demand, which is fundamental to the reliability of the system. 

In the scenario 2030a it is assumed that only Wind and Solar will expand in order to supply the additional 137 TWh needed, which results in a very aggressive scenario of development for these two sources. Scenario 2030b is based on an expansion of the same two sources but includes also aggressive conservation of electricity by widely deploying solar water heaters (SWH) in order to replace electric showerheads \cite{Giglio2016,Giglio2019,Naspolini2019,Cruz2020}.

To develop the 2030b scenario first we searched for a good estimate of the total energy consumption of showerheads in Brazil. Corrêa da Silva et al. \cite{CorreaDaSilva2016a} have considered a value of 20 TWh/year in 2004, while the most recent estimate by EPE, previously mentioned in this work, was about 31 TWh/year for 2017, including losses. In the higher hand, Cruz et al. \cite{Cruz2020} have estimated that in 2020 this amount of energy should be near 55 TWh/year. On the other hand, Cardemil et al. \cite{Cardemil2018} have estimated the showerhead load profile of an average day by considering an annual consumption of 33.7 TWh, including losses for the year 2012. In this way, we found reasonable to estimate potential savings due to SWH at about 50 TWh per year in 2030. 

\begin{figure}[!h]
\centering
\includegraphics[scale=0.3]{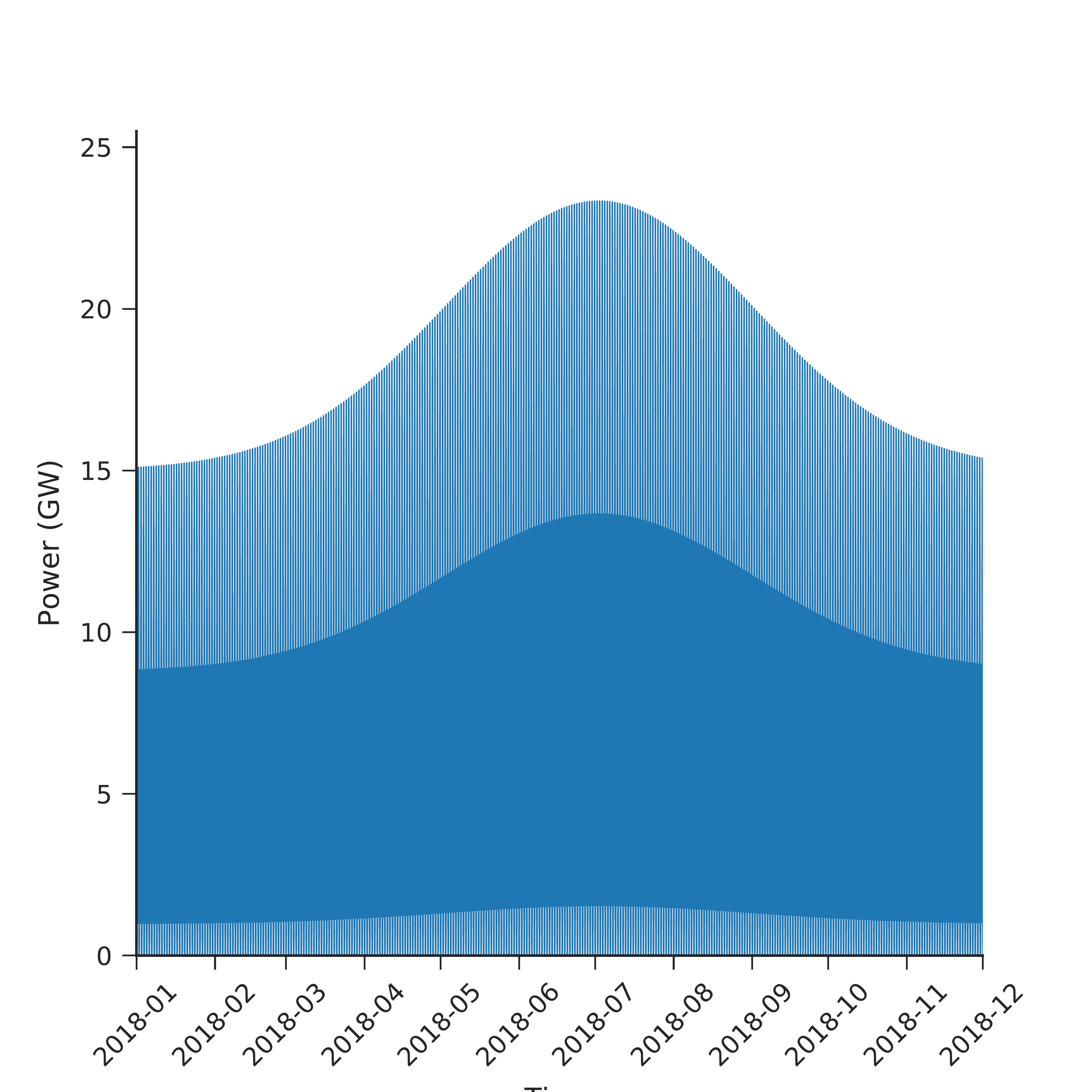}%
\caption{Electric showerhead consumption estimated for an entire year in Brazil, with hourly resolution.}
\label{shower}
\end{figure}

Next, we looked for an estimation of the hourly consumption of electricity due to electric showerheads. Based on localized measurements found in the literature \cite{Cardemil2018,Naspolini2017a} we choose to distribute the daily consumption using three Gaussian curves centered at 7, 12, and 19 hours. The result is described by equation \ref{eq1},
\begin{equation}
P_{sh}(t)=\frac{1}{2 \sigma}\left(35 e^{-1/2(\frac{t-7}{\sigma})^2}+5e^{-1/2(\frac{t-12}{\sigma})^2}+60e^{-1/2(\frac{t-19}{\sigma})^2}\right),
\label{eq1}
\end{equation}
where each Gaussian has a weight factor of 35, 5, and 60 percent, while $\sigma=1.8$ was used. This value was chosen in order to have the shape of the showerhead daily load similar to literature data. Though, another Gaussian was used to distribute the amplitude of the consumption throughout the year, centering the Gaussian in the middle of it, once the showerhead consumption is more significant in the winter. The result is the profile shown in figure \ref{shower}.

In the scenario 2030c the additional 137 TWh are supplied by Wind, Nuclear, and Solar, while in the last scenario, 2030d, with the goal to achieve enhanced reliability of the national supply it is assumed a reduction in total annual hydro generation, expansion of all other sources and conservation due to SWH. It is important to explain that though the nominal hydropower generation capacity in Brazil is huge and about 187 TWh of electricity can be stored in the reservoirs over the country, as the water level decreases the nominal capacity is lost \cite{Hunt.2018}. In table \ref{tab1} we show the details of the four scenarios in comparison to the 2018 data.
\begin{table}[h]
\centering
\begin{tabular}{|c|c|c|c|c|c|}
\hline
\textbf{Source}&\textbf{2018}&\textbf{2030a}&\textbf{2030b}&\textbf{2030c}&\textbf{2030d}\\
\hline
Hydro&407.0& & & & (0.8$\times$)325.6\\
Thermal&80.4& & & &(1.5$\times$)120.6\\
Wind&46.0&(2.5$\times$)115.0 &(2$\times$)92.0 &(2$\times$)92.0 &(2$\times$)91.9\\
Nuclear&15.7& & &(4.1$\times$)64.2 &(1.7$\times$)26.6\\
Solar&2.7&(26$\times$)69.6 & (16$\times$)42.9&(16$\times$)42.9 &(26$\times$)69.6   \\
SWH&-&-&50.5&- & 50.5 \\
\hline
Total& 551.8&687.7&688.5 &686.5 &684.8 \\
\hline
\end{tabular}
\caption{Production of electricity (in TWh) by source in 2018 and in the four different projections made for 2030.}
\label{tab1}
\end{table}

By taking the hourly production curve in 2018 for Thermal, Wind, Nuclear, and Solar, as well the model for conservation of electricity by SWH, we calculated what it would be the remaining power left to be supplied by hydro. To remove any ``noise'', clearing the visualization of trends, we fitted the data for each year with a Generalized Additive Model (GAM) \cite{Hill1986}. This process was performed by using the PyGam library, in Python. A Tensor term was used with the day of the week and year fraction as features, with 200 and 20 splines, respectively. The fitting obtained had a pseudo $r^2$ higher than 0.9 in all cases. In the next section, we analyze the hourly load and compare the results obtained in each of our scenarios.

\section{Hourly load analysis}

\begin{figure}[!h]
\centering
\includegraphics[scale=1.2,trim={1cm 10cm 12cm 0cm},clip]{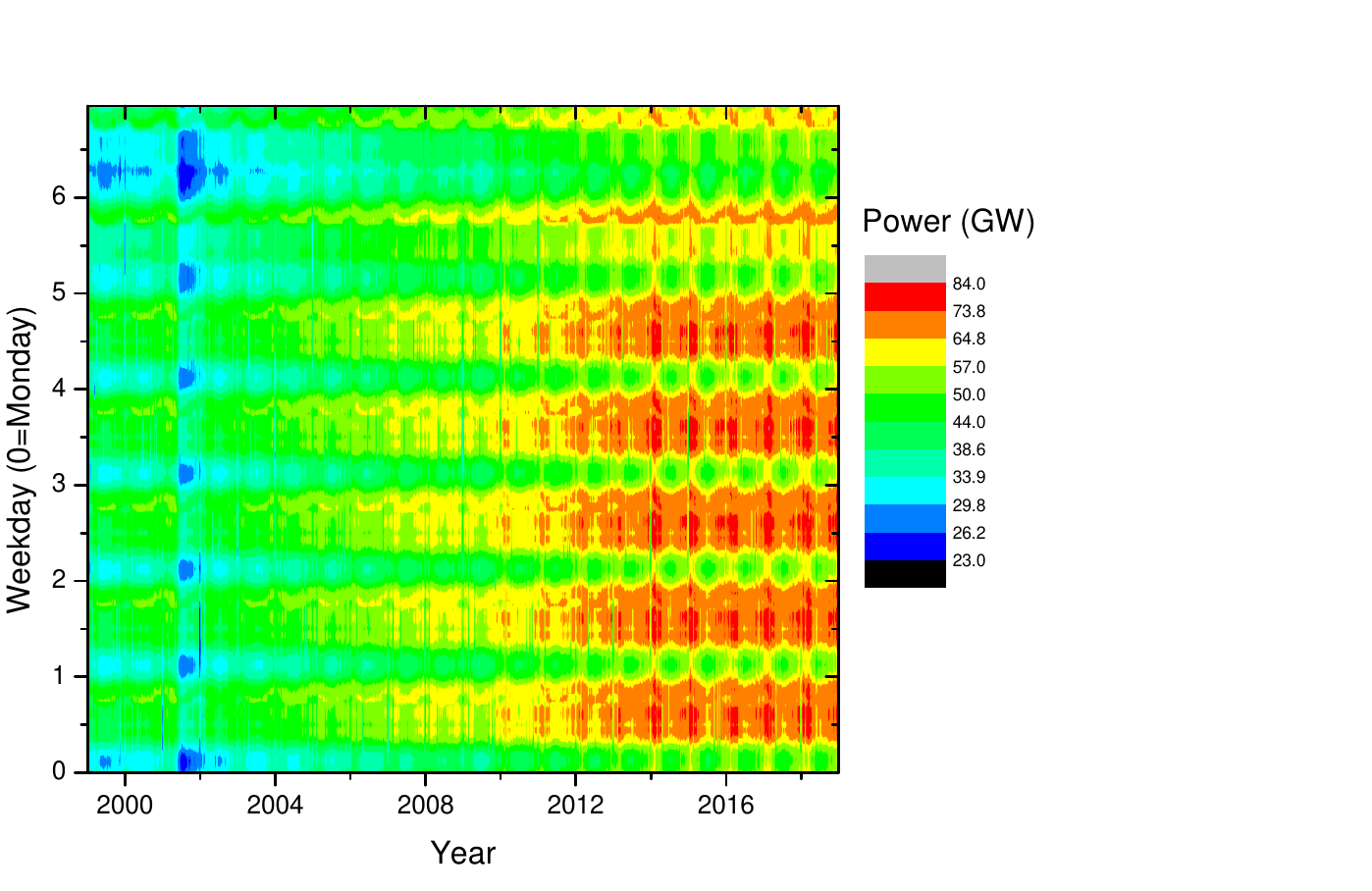}%
\caption{Evolution of power consumption in Brazil with hourly resolution.}
\label{figbruta}
\end{figure}

\begin{figure}[!h]
\centering
\includegraphics[scale=1.2,trim={1cm 10cm 12cm 0cm},clip]{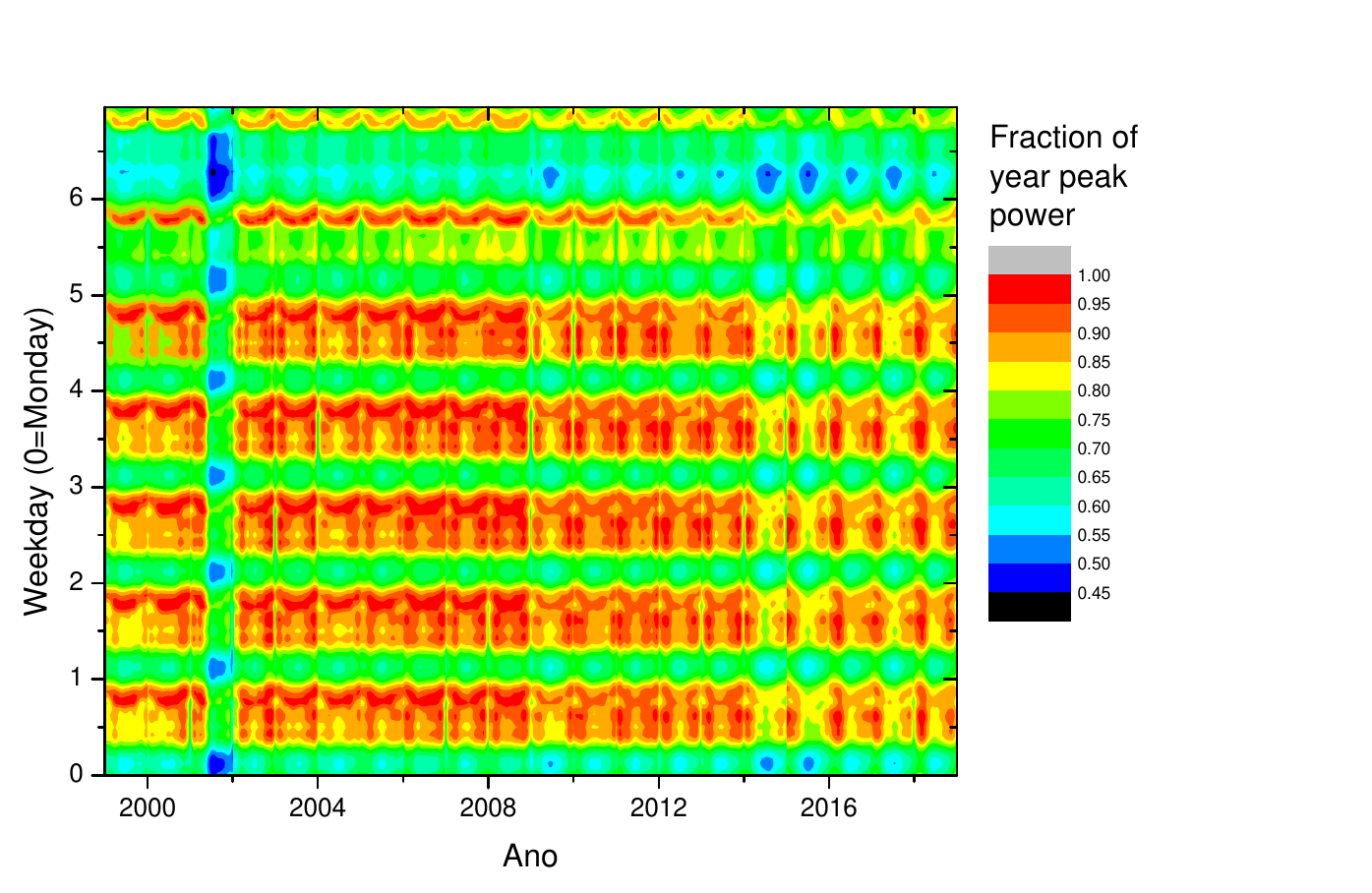}%
\caption{Evolution of power consumption as fraction of annual peak power.}
\label{fignorma}
\end{figure}

In figure \ref{figbruta} it is shown a contour plot of load power in the Brazilian national grid against weekday and year, with hourly resolution.
The well-marked load reduction in the second half of 2001 happened when hydropower supplied more than 90\% of Brazil's electricity needs, and it is well known as a drought triggered crisis \cite{Jardini2002} as we mentioned before. As the total consumption per year has increased by about 60\% in the period shown in this figure, it is quite difficult to observe effects due to the 2014 crisis. In this way, to remove ``noise'' and better identify the trends, we have fitted the data for each year with a GAM and normalized the power by the maximum load in that year. These results are shown in figure \ref{fignorma}.

As one can see, the load pattern over a week has changed significantly during the last 20 years. In the early 2000s, the load peaks were found exclusively in evenings and included a seasonal dependence. One possible explanation for this trend is the widespread use of showerheads, which are by far the first source of heat for the shower in Brazil and consume a significant amount of electricity in a quite narrow time range \cite{Cruz2020}. 

After 2001 the country experienced a period of strong economic growth and except by the year 2009, when the country was affected by the global financial crisis, the electricity consumption increased by about 4\% every year until 2014. As one can see in figure \ref{fignorma} the load demand pattern has changed from early in the night to summers (February/March), mostly in the afternoon. This can be related to the increased share of the services sector in the Brazilian economy, as well as the widespread adoption of air conditioning systems. On the other hand, blue regions in figure \ref{fignorma} denote loads near 50\% of the year's peak and may indicate a relative reduction in industrial activity. In this sense, dark blue regions are observed in late evenings trough out the second half of 2001, as well on Mondays and Sundays of periods with constrained economic activity (2009, 2014 and 2015).

\subsection{2030 Scenarios} 

\begin{figure}[!h]
\centering
\includegraphics[scale=1.3,clip,trim={1cm 11cm 12cm 0cm}]{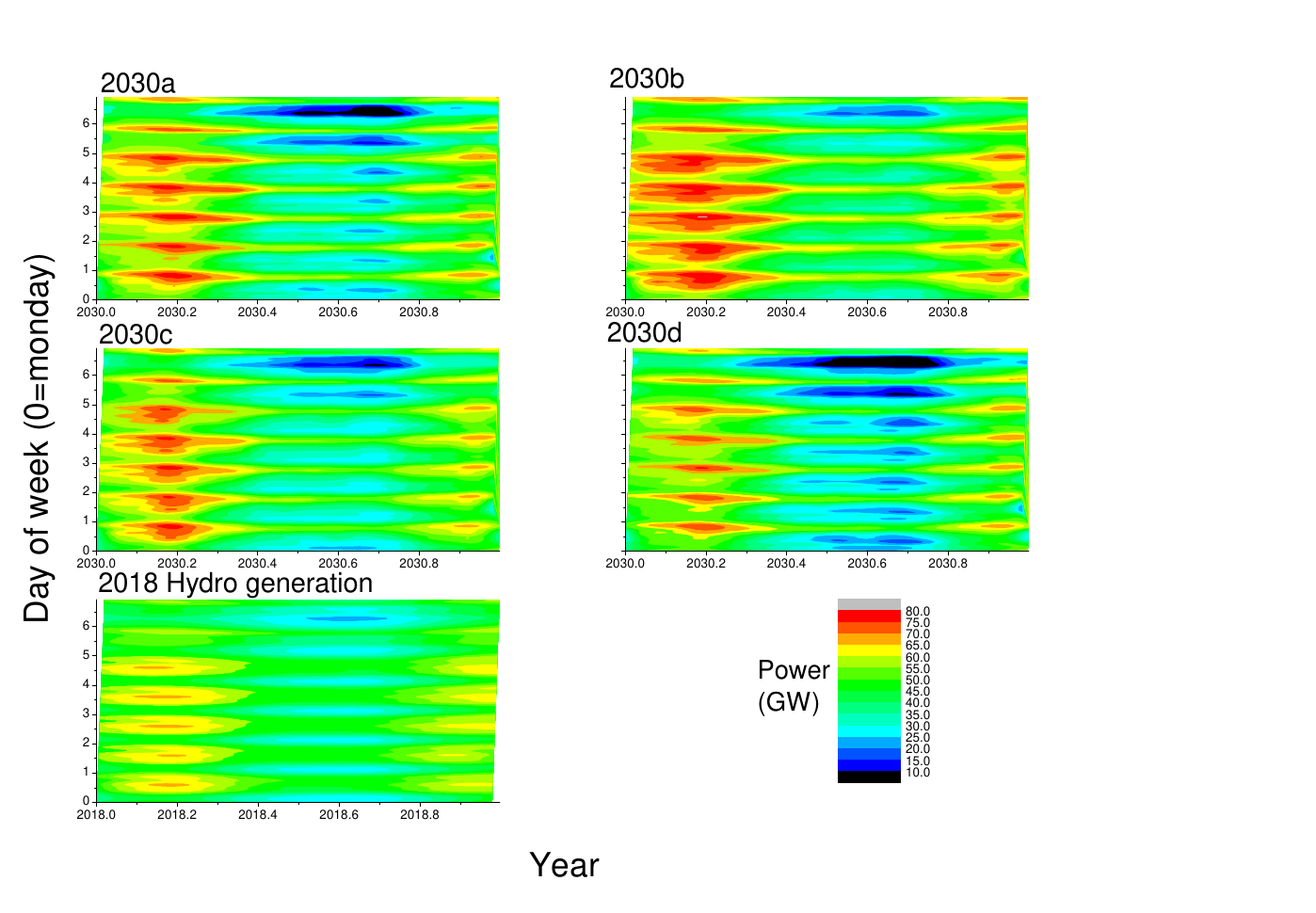}%
\caption{Simulated hydro power demand for 2030 scenarios and 2018 hydro generation. The scenarios 2030a, 2030b and 2030c have exactly the same amount of hydro power generated in 2018, while 2030d has 80\%.}
\label{figmapa}
\end{figure}

The four scenarios mentioned in table \ref{tab1} were evaluated assuming the 2018 generation load for each source, multiplied by a factor, as indicated in the table \ref{tab1}. Next the generation from wind, solar, thermal, nuclear and SWH were subtracted from the demand load to estimate what is the load profile left, which should be fulfilled by hydro power. The results are shown in figure \ref{figmapa}.

The contour plot should be carefully interpreted. These are the plots of the GAM obtained by fitting the data. The red color, for example, represents a higher chance to have this hydropower demand. So, one may say that in scenario 2030a the peaks are more concentrated in early evenings during summer, while in 2030b it is more distributed in the afternoon once this scenario has lower solar photovoltaics, which is likely to peak in this period.

Brazil has today about 85 GW of hydroelectric dams and 12 GW of run-of-river hydropower \cite{Dranka2018a}. Even though our scenarios do not consider an increase in annual hydropower production, in all of them hydro needs to provide higher power for more time, mainly at the end of the summer (February/March). For example, in 2018 the Hydropower generation in Brazil delivered more than 70 GW for only 12 hours, while this number reaches 429, 289 and 360 in the scenarios 2030a, 2030b and 2030c, respectively. This result indicates an increased risk for the national grid, once the maximum hydropower available is often constrained due head loss in the reservoirs, which has already played a central role in the last energy crisis in the country. As Hunt et al \cite{Hunt.2018} have discussed, in 2012 the operator of the national grid have taken some ``optimistic decisions", like ``to spill some of the water in the Furnas Reservoir wasting hydroelectric potential, to increase peak generation in the Grande and Paraná Rivers, and reduce thermoelectric generation". This kind of management has worsened the energy crisis experienced in 2014-2015. In the scenario 2030d hydropower should supply more than 70 GW by only 10 hours. In such projection hydropower would supply only 80\% of what is has generated in 2018, and the results indicate that such aggressive conservation should be necessary in order to keep about the same level of power demand from hydro experienced in 2018.

\begin{figure}[!h]
\centering
\includegraphics[scale=0.25,clip]{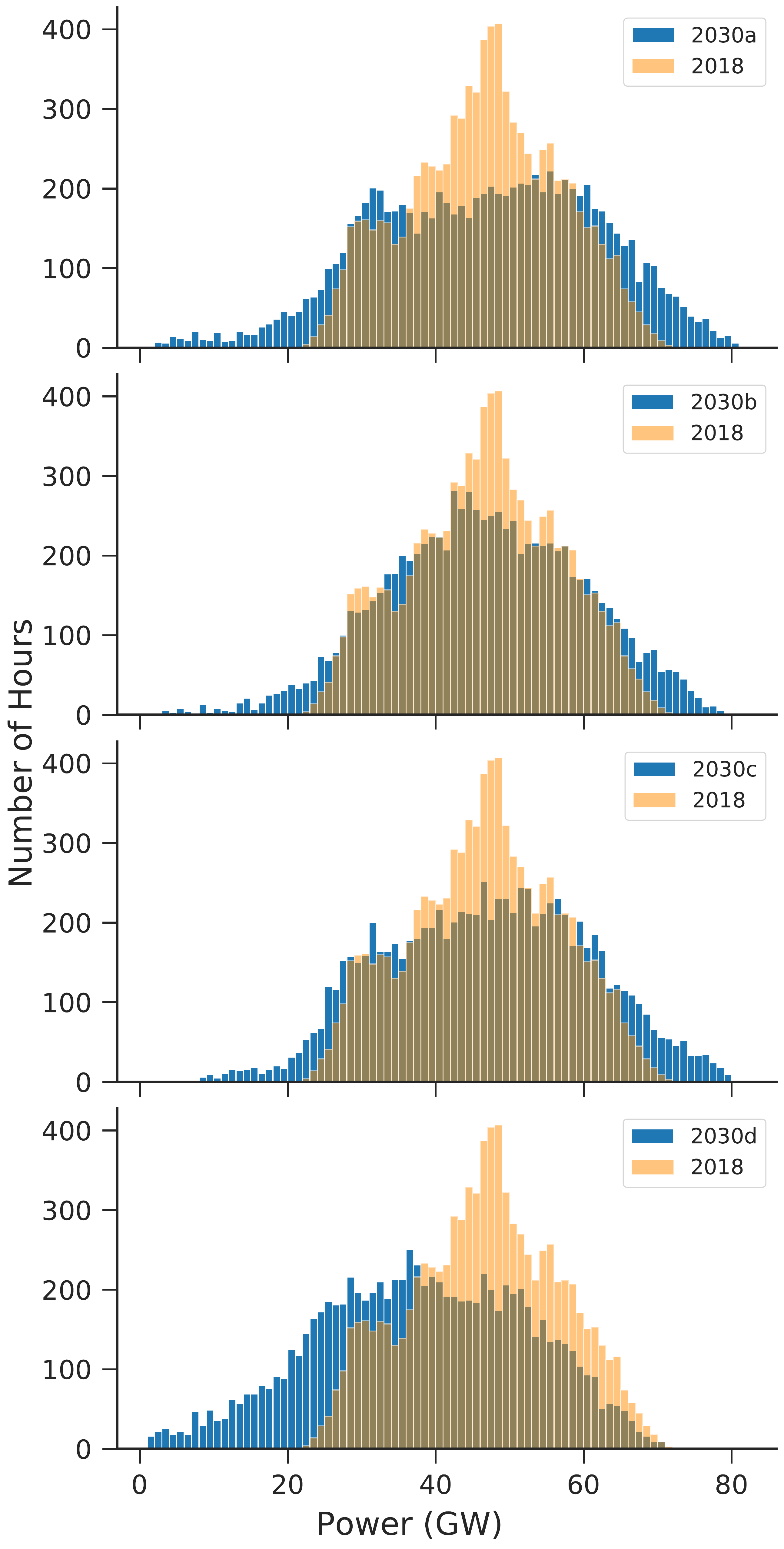}%
\includegraphics[scale=0.25,clip]{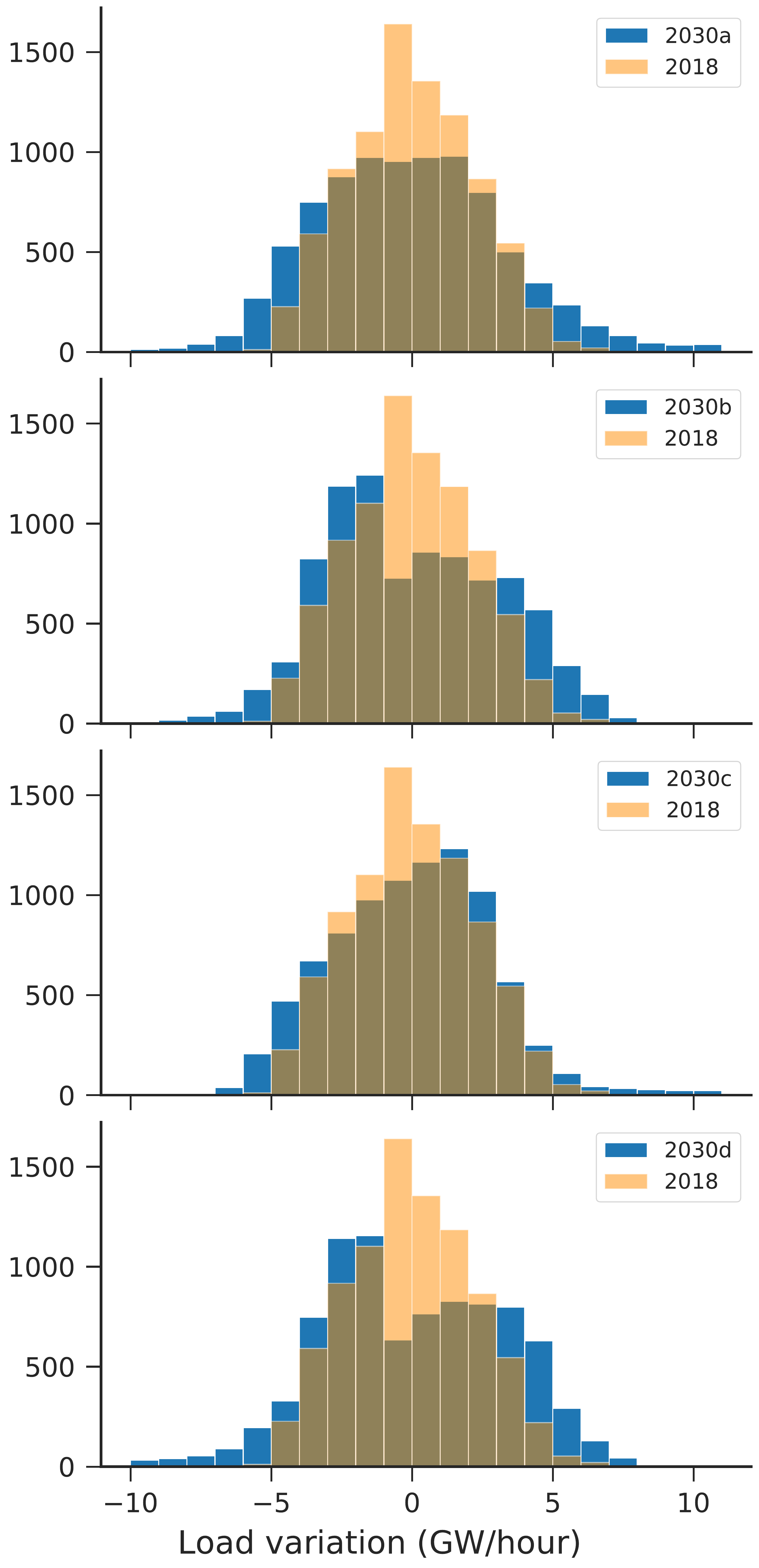}%
\caption{Histograms of load (left column) and load variation (right column) to be supplied by hydro in our scenarios compared to 2018 data.}
\label{fighisto}
\end{figure}

On other hand, during winters hydropower could operate most of the time at reduced power compared to 2018, though another problem emerges as reduce hydropower below 25 GW may not be practical because some minimum water flow is required. Additionally one may expect that higher penetration of intermittent power sources will increase the rate at which hydropower increases and decreases. To quantify the trends mentioned above we show in figure \ref{fighisto} histograms of expected power from hydro (number of hours operating at each power) and power variation (number of times that each variation in hydropower would be required) for our scenarios compared to 2018 data.

As one can see, in all scenarios there is a significant shift in the histograms. In the first three of them, as we mentioned before, the number of hours in which hydropower operates above its 2018 maximum (70 GW) increases significantly. On the other hand the number of hours operating below the 2018 minimum hydropower (20 GW) also increases. At right in figure \ref{fighisto} one can see that in 2018 the hydropower production barely oscillated more than 5 GW in one hour, while in the scenario 2030a it could be required sometimes to increase this production by 11 GW/hour. The scenario less aggressive concerning such ramps is the 2030c, where we have less solar and wind and more nuclear than in the scenario 2030a.

\section{Discussion}

As Brazil does not have any gigawatt-scale hydropower plant in the construction phase, we found it reasonable to evaluate scenarios, where investment will be made in other sources, such as Wind, Solar, Nuclear Thermal, or electricity consumption is avoided due conservation introduced by solar water heaters. When hourly load and production by these sources are taken into account, the outcome achieved in each scenario and its impact in the management of hydropower could be evaluated. 

In 2018 the highest production of hydropower was concentrated in afternoons between February and March. A massive investment in solar power resulting in 26 times more peak power than the 2018 level (2030a and 2030d) would result in shifting hydropower peak for early evenings, and the lowest hydropower demand should be experienced during winters, on Saturday and Sunday mornings, at levels that may not be practical. Often the precipitation levels in many regions may be quite high until June or July, in such a way that this time is likely to be the period where reservoirs are full. In other words, they would be full when lower power is demanded, increasing the risk to have water spilled and to have a significant head loss of the reservoirs in the following summer. 

More conservative scenarios such as 2030b and 2030c, where solar power is increased by a factor of 16 would make hydro to operate at its highest power levels in afternoons and evenings, but the effect on Saturday and Sunday mornings during winter are attenuated. It is especially interesting to compare these two scenarios once both have the same amount of Wind and Solar, but the first uses SWH while the second considers an increase in nuclear power. 

Martins et al. \cite{Martins2012} have investigated also the financial and economical aspects of SWH and payback time for a small scale system of only 4 years was obtained, which is compatible with the value estimated by Naspolini et al. \cite{Naspolini2019}. On the other hand, Cruz et al. \cite{Cruz2020} have calculated that for about 17.9\% of the Brazilian households it does make sense from the economical point of view to invest in SWH, while only 6\% of the residences have an SWH system in 2018. However, such economic benefits are often underestimated because in practice the price of electricity at peak hours is subsidized in Brazil \cite{Naspolini2016,Naspolini2019}. In this context SWH could be an important alternative to reduce the demand of electricity and reduce the need to expand thermal or nuclear power.

Finally one may conclude that increasing the share of intermittent renewables in the grid are very likely to affect dramatically the hydropower system in Brazil. An increase in the peak hydropower, longer periods demanding more than 70 GW or less than 25 GW of hydropower are very likely, and it seems likely to impose a challenge to the Brazilian national grid.  To enhance the reliability of the electricity supply without increasing the storage capacity seems not possible, though we estimated that hydropower can keep filling the gap between supply and demand until certain increased penetration of Solar and Wind. Among the technologies readily available for large scale deployment, SWH seems a feasible one that could significantly contribute to the reliability of the Brazilian national grid.

\bibliographystyle{model1-num-names}
\bibliography{library}
\end{document}